\documentclass[conference,10pt]{IEEEtran}
\IEEEoverridecommandlockouts
\pdfoutput=1 %
\usepackage{cite}
\usepackage{amsmath,amssymb,amsfonts,bm}
\usepackage{graphicx}
\usepackage{textcomp}
\usepackage{xcolor}
\usepackage[caption=false,font=footnotesize]{subfig}
\usepackage{xfrac} %
\usepackage{glossaries} %
\usepackage{tikz,pgfplots}
\usetikzlibrary{fit,calc,external}

\pgfplotsset{compat=1.16}
\definecolor{darkblue}{rgb}{0,0.2706,0.541}
\definecolor{darkgreen}{rgb}{0.098,0.4784,0.5176}
\definecolor{lightgreen}{rgb}{0.3412,0.7411,0.7647}
\definecolor{orange}{rgb}{0.9255,0.4313,0}
\definecolor{darkred}{rgb}{0.7529,0,0}
\definecolor{black}{rgb}{0,0,0}

\def\BibTeX{{\rm B\kern-.05em{\sc i\kern-.025em b}\kern-.08em
    T\kern-.1667em\lower.7ex\hbox{E}\kern-.125emX}}

\newcommand{\mvec}[1]{\textbf{#1}}%
\newcommand{\mvecm}[1]{\bm{#1}}%
\newcommand{\mset}[1]{\mathcal{#1}}
\newcommand{\sgn}[1]{\text{sgn}(#1)}

\makeatletter
\def\ps@IEEEtitlepagestyle{%
	\def\@oddfoot{\mycopyrightnotice}%
	\def\@evenfoot{}%
}
\def\mycopyrightnotice{%
	{\footnotesize This work has been submitted to the IEEE GLOBECOM 2024 for possible publication.\hfill}%
	\gdef\mycopyrightnotice{}%
}

\begin{document}

\newacronym{sc}{SC}{successive cancellation}
\newacronym{ssc}{SSC}{simplified successive cancellation}
\newacronym{fssc}{FSC}{fast successive cancellation}
\newacronym{scl}{SCL}{successive cancellation list}
\newacronym{ca-scl}{CA-SCL}{CRC-aided \gls{scl}}
\newacronym{llr}{LLR}{log-likelihood ratio}
\newacronym{lut}{LUT}{lookup table}
\newacronym{bpsk}{BPSK}{binary phase-shift keying}
\newacronym{awgn}{AWGN}{additive white Gaussian noise}
\newacronym{fer}{FER}{frame error rate}
\newacronym{bler}{BlER}{block error rate}
\newacronym{spc}{SPC}{single parity check}
\newacronym{ib}{IB}{information bottleneck}
\newacronym{crc}{CRC}{cyclic redundancy check}
\newacronym{noder0}{$R0$}{rate-0}
\newacronym{noder1}{$R1$}{rate-1}
\newacronym{noderep}{$Rep$}{repetition}
\newacronym{nodespc}{$SPC$}{single parity check}
\newacronym{ml}{ML}{Most-Likely}

\title{Finite Alphabet Fast List Decoders for Polar Codes
}

\author{\IEEEauthorblockN{Syed Aizaz Ali Shah and Gerhard Bauch}
\IEEEauthorblockA{\textit{Institute of Communications}, \textit{Hamburg University of Technology}, Hamburg, Germany \\
\{aizaz.shah; bauch\}@tuhh.de}\\[-0.6cm]
}

\maketitle

\begin{abstract}
The so-called fast polar decoding schedules are meant to improve the decoding speed of the sequential-natured successive cancellation list decoders. 
The decoding speedup is achieved by replacing various parts of the serial decoding process with efficient special-purpose decoder nodes.
This work incorporates the fast decoding schedules for polar codes into their quantized finite alphabet decoding. In a finite alphabet successive cancellation list decoder, the \glsentrylong{llr} computations are replaced  with lookup operations on low-resolution integer messages. The lookup tables are designed using the information bottleneck method. It is shown that the finite alphabet decoders can also leverage  the special decoder nodes found in the literature. Besides their inherent decoding speed improvement, the use of these special decoder nodes drastically reduces the number of lookup tables required to perform the finite alphabet decoding. 
In order to perform quantized decoding using lookup operations, the proposed decoders require up to $93\%$ less unique lookup tables as compared to the ones that use the conventional successive cancellation schedule.
Moreover, the proposed decoders exhibit negligible loss in error correction performance without necessitating alterations to the lookup table design process.
\end{abstract}

\section{Introduction}

Besides error correction, decoding speed is also an important factor in error control.
The sequential nature of the \gls{sc}\cite{arikan_channel_2009} decoding  hampers the decoding speed of polar codes. In order to tackle this issue, ways to make the \gls{sc} decoding faster have been discovered besides exploring other decoding schemes, e.g., belief propagation. 
In that regard, various steps in the \gls{sc} schedule have been identified where a block of bits is decoded in one shot using an efficient constituent decoder node  instead of the serial decoding.
First, \gls{ssc} decoder was proposed where \gls{noder0}  and \gls{noder1} constituent decoders were exploited \cite{Alamdar_SSC_2011}. 
Later, \gls{noderep} and \gls{nodespc} constituent decoders were utilized in \gls{sc} decoding to achieve faster decoding\cite{Sarkis_FastSSC_2014}. 
The use of \gls{noder0}, \gls{noder1}, \gls{noderep} and \gls{nodespc} nodes was also extended to \gls{scl}\cite{Sarkis_2016_FSSCL,Hashemi2016_SSCL_SCL_equivalence,Hashemi2017_FSSL_list_size}.  Other special constituent nodes and their decoding  can be found in the literature, e.g., \cite{Sarkis_FastSSC_2014,Hanif_2017_other_nodes,Condo_2018_other_nodes,Shen_2022_SeqRep}. 
An \gls{sc} or \gls{scl} decoder that exploits such special nodes is referred to as a \textit{fast} decoder. Here, the term \gls{fssc} decoder is used for a decoder that makes use of \gls{noder0}, \gls{noder1}, \gls{noderep}  and \gls{nodespc} nodes.

The resolution of the reliability messages exchanged in the decoding process plays a significant role in the efficient hardware implementation of an \gls{scl} decoder.
Ideally,  a small bit-width  with acceptable degradation in the error correction performance of the decoder is used. 
One way to navigate this trade-off is the finite alphabet decoding paradigm  where $w$-bit integer-valued messages communicate reliability information.
In \cite{shah_design_2019,shah_coarsely_2019,Koike-Akino2019}, the \gls{ib} method was used to design finite alphabet \gls{sc} and \gls{scl} decoders where the decoding operations are realized as mutual information maximizing \glspl{lut}.  
Two types of \glspl{lut} are used decoding  an $N$-bit codeword: $2N{-}2$ \textit{decoding} tables that replace the \gls{llr} computations with lookup operations of integers. $N$ \textit{translation} tables that translate the integer-valued messages into \glspl{llr} for path metric updates in the list decoding.
It was shown that a $64$-bit \gls{scl} decoder outperfomrs  a 4-bit quantized \gls{ib} \gls{scl} decoder by only a small margin\cite{shah_MSIB_2023}.

The use of \glspl{lut} designed with the \gls{ib} method was recently combined with the \gls{ssc} decoding in \cite{GiardShah2023}. This work extends the use of mutual information maximizing \glspl{lut} to \gls{fssc} list decoding. 
It is shown that the \glspl{lut} designed for \gls{sc} schedule\cite{shah_design_2019,shah_MSIB_2023} are readily usable with the \gls{fssc} schedule where efficient decoders for the special nodes from the literature\cite{Hashemi2017_FSSL_list_size} are used. 
Moreover, the proposed finite alphabet decoders require a considerably smaller number of decoding and translation tables.
The fast decoding schedule has negligible effect on the error correction performance of an \glspl{lut}-based \gls{scl} decoder.

\section{Polar Codes Review}

\subsection{Polar Codes}
A polar code with length $N=2^n$, where $n=1,2,\ldots$, is described by its $N\times N$ generator matrix $\mvec{F}^{\otimes n}$ where ${\otimes n}$ represents the $n$th Kronecker product
with $\mvec{F}^{\otimes 1}=\mathbf{F}{=}\left[\begin{smallmatrix}
    1	&	1 \\
    0	&	1
\end{smallmatrix}\right]$\cite{arikan_channel_2009}.
For a code rate of $R{=}\sfrac{K}{N}$, $N{-}K$ bits in $\textbf{u}=[u_0, \dots u_{N-1}]^T$ are set to $0$ in this work, and referred to as the \textit{frozen} bits.
The values and locations of the frozen bits are known to the decoder. The remaining  $K$ positions in \textbf{u}, specified in the \textit{information} set $\mset{A}$, carry the information bits.
The encoding  follows as $\textbf{x} {=} \mvec{F}^{\otimes n} \mvec{u}$.

\begin{figure}[t]
\centering
  \begin{minipage}{0.6\columnwidth}
    \centering
    \newcommand{\ubit}[1]{$u_{#1}$}
\newcommand{\fbit}[1]{\color{gray}$u_{#1}$}
\newcommand{\ucw}[1]{$x_{#1}$}
\newcommand{\fcw}[1]{\color{gray}$x_{#1}$}
\newcommand{\ub}[1]{$#1$}
\newcommand{\fb}[1]{\color{gray}$#1$}

\begin{tikzpicture}

\usetikzlibrary{shapes,positioning,arrows,decorations.markings,fit}

\definecolor{varnode_fill}{RGB}{0,0,0}
\definecolor{chknode_fill}{RGB}{255,255,255}

\tikzset{
  chknode/.style={draw,fill=chknode_fill,circle,minimum size=0.3cm, inner sep=0},
  varnode/.style={draw,fill=varnode_fill,circle,minimum size=0.1cm, inner sep=0},
  channel/.style={draw,fill=white,rectangle},
  sep/.style={rectangle,minimum width=0.25cm, inner sep=0},
  empty/.style={rectangle, inner sep=0},
  bit/.style={circle, inner sep = 0}
}

\tikzset{green dotted/.style={draw=green!50!black, line width=1pt,
    dash pattern=on 3pt off 3pt,
    inner sep=0.4mm, rectangle, rounded corners}};

\matrix[row sep=2mm, column sep=1.8mm] {
  \node[bit] (n0s0) {\ub{u_0}}; & \node[chknode] (n0s1) {$+$}; & \node[sep] (s10) {}; & \node[chknode] (n0s2) {$+$}; & \node[empty] {};              & \node[sep] (s20) {}; & \node[chknode] (n0s3) {$+$}; & \node[empty] {}; & \node[empty] {}; & \node[empty] {}; && \node[bit] (xn0s4) {\ub{x_0}};\\
  \node[bit] (n1s0) {\ub{u_1}}; & \node[varnode] (n1s1) {};    & \node[sep] (s11) {}; &                              & \node[chknode] (n1s2) {$+$};  & \node[sep] (s21) {}; & \node[empty] {};             & \node[chknode] (n1s3) {$+$}; & \node[empty] {}; & \node[empty] {}; && \node[bit] (xn1s4) {\ub{x_1}};\\
  \node[bit] (n2s0) {\ub{u_2}}; & \node[chknode] (n2s1) {$+$}; & \node[sep] (s12) {}; & \node[varnode] (n2s2) {};    & \node[empty] {};              & \node[sep] (s22) {}; & \node[empty] {};             & \node[empty] {}; & \node[chknode] (n2s3) {$+$}; & \node[empty] {}; && \node[bit] (xn2s4) {\ub{x_2}};\\

  \node[bit] (n3s0) {\ub{u_3}}; & \node[varnode] (n3s1) {};    & \node[sep] (s13) {}; & \node[empty] {};             & \node[varnode] (n3s2) {};     & \node[sep] (s23) {}; & \node[empty] {};             & \node[empty] {}; & \node[empty] {}; & \node[chknode] (n3s3) {$+$}; && \node[bit] (xn3s4) {\ub{x_3}};\\

  \node[bit] (n4s0) {\ub{u_4}}; & \node[chknode] (n4s1) {$+$}; & \node[sep] (s14) {}; & \node[chknode] (n4s2) {$+$}; & \node[empty] {};              & \node[sep] (s24) {}; & \node[varnode] (n4s3) {};    & \node[empty] {}; & \node[empty] {}; & \node[empty] {}; && \node[bit] (xn4s4) {\ub{x_4}};\\
  \node[bit] (n5s0) {\ub{u_5}}; & \node[varnode] (n5s1) {};    & \node[sep] (s15) {}; &                              & \node[chknode] (n5s2) {$+$};  & \node[sep] (s25) {}; & \node[empty] {};             & \node[varnode] (n5s3) {}; & \node[empty] {}; &  \node[empty] {}; && \node[bit] (xn5s4) {\ub{x_5}};\\
  \node[bit] (n6s0) {\ub{u_6}}; & \node[chknode] (n6s1) {$+$}; & \node[sep] (s16) {}; & \node[varnode] (n6s2) {};    & \node[empty] {};              & \node[sep] (s26) {}; & \node[empty] {};             & \node[empty] {}; & \node[varnode] (n6s3) {}; &  \node[empty] {}; && \node[bit] (xn6s4) {\ub{x_6}};\\
  
  \node[bit] (n7s0) {\ub{u_7}}; & \node[varnode] (n7s1) {};    & \node[sep] (s17) {}; &                              & \node[varnode] (n7s2) {};  & \node[sep] (s27) {}; & \node[empty] {};             & \node[empty] {}; & \node[empty] {}; &  \node[varnode] (n7s3) {}; && \node[bit] (xn7s4) {\ub{x_7}};\\
};
\path[-] (n0s0) edge (n0s1) (n0s1) edge (n0s2) (n0s2) edge (n0s3) (n0s3) edge (xn0s4);
\path[-] (n1s0) edge (n1s1) (n1s1) edge (n1s2) (n1s2) edge (n1s3) (n1s3) edge (xn1s4);
\path[-] (n2s0) edge (n2s1) (n2s1) edge (n2s2) (n2s2) edge (n2s3) (n2s3) edge (xn2s4);
\path[-] (n3s0) edge (n3s1) (n3s1) edge (n3s2) (n3s2) edge (n3s3) (n3s3) edge (xn3s4);
\path[-] (n4s0) edge (n4s1) (n4s1) edge (n4s2) (n4s2) edge (n4s3) (n4s3) edge (xn4s4);
\path[-] (n5s0) edge (n5s1) (n5s1) edge (n5s2) (n5s2) edge (n5s3) (n5s3) edge (xn5s4);
\path[-] (n6s0) edge (n6s1) (n6s1) edge (n6s2) (n6s2) edge (n6s3) (n6s3) edge (xn6s4);
\path[-] (n7s0) edge (n7s1) (n7s1) edge (n7s2) (n7s2) edge (n7s3) (n7s3) edge (xn7s4);

\path[-] (n0s1) edge (n1s1);
\path[-] (n2s1) edge (n3s1);
\path[-] (n4s1) edge (n5s1);
\path[-] (n6s1) edge (n7s1);

\path[-] (n0s2) edge (n2s2);
\path[-] (n1s2) edge (n3s2);
\path[-] (n4s2) edge (n6s2);
\path[-] (n5s2) edge (n7s2);

\path[-] (n0s3) edge (n4s3);
\path[-] (n1s3) edge (n5s3);
\path[-] (n2s3) edge (n6s3);
\path[-] (n3s3) edge (n7s3);

\path[-,darkred,line width=1.5pt] ($(n0s2.east)+(0.7,0.0)$) edge (n0s3) (n0s3) edge (xn0s4);
\path[-,darkred,line width=1.5pt] (n0s3) edge (n4s3);
\path[-,darkred,line width=1.5pt] ($(n4s2.east)+(0.7,0.0)$) edge (n4s3) (n4s3) edge (xn4s4);

\node (b0) at ($(n0s2.east)+(0.8,-0.20)$) {\color{gray}$b_0$};
\node (b0) at ($(n4s2.east)+(0.8,-0.20)$) {\color{gray}$b_4$};

\tikzset{Nv_8/.style={draw=darkgreen, line width=1pt,
    dash pattern=on 3pt off 3pt,
    inner sep=0.4mm, rectangle, rounded corners}};
\tikzset{Nv_4/.style={draw=orange, line width=1pt,
    dash pattern=on 3pt off 3pt,
    inner sep=0.4mm, rectangle, rounded corners}};
\tikzset{Nv_2/.style={draw=darkblue, line width=1pt,
    dash pattern=on 3pt off 3pt,
    inner sep=0.4mm, rectangle, rounded corners}};
\node (g_N8) [Nv_8, fit = (n0s3) (n3s3) (n4s3) (n7s3)] {};
\node (g_N4_0) [Nv_4,fill, fill opacity =0.2, fit = (n0s2) (n1s2) (n2s2) (n3s2)] {};
\node (g_N4_1) [Nv_4, fit = (n4s2) (n5s2) (n6s2) (n7s2)] {};
\node (g_N2_0) [Nv_2, fit = (n0s1) (n1s1)] {};
\node (g_N2_1) [Nv_2, fit = (n2s1) (n3s1)] {};
\node (g_N2_2) [Nv_2, fit = (n4s1) (n5s1)] {};
\node (g_N2_3) [Nv_2, fit = (n6s1) (n7s1)] {};

\node (d1) at ($(g_N8.north)+(0.0,0.20)$) {\color{darkgreen}$d=1$};
\node (d1) at ($(g_N4_0.north)+(0.0,0.20)$) {\color{orange}$d=2$};
\node (d1) at ($(g_N2_0.north)+(0.0,0.20)$) {\color{darkblue}$d{=}3$};
\node (v) at ($(n3s2.west)+(-0.3,0.25)$) {$v$};
\end{tikzpicture}
  \end{minipage}%
  \vspace{-10pt}
  \caption{Graph representation of a polar code with $N=8$.}
  \label{fig:pc8}
 \vspace{-10pt}
\end{figure}
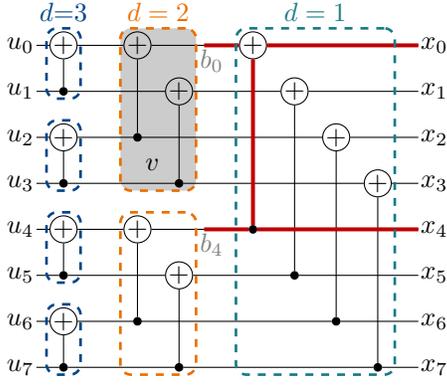

A polar code can be  represented as a graph like that of Fig.\,\ref{fig:pc8} for $N{=}8$, where $\oplus$ represents modulo-2 addition (XOR). 
For encoding with a given $\mset{A}$, a codeword $\mvec{x}$ is generated by propagating the frozen and information bits in $\mvec{u}$ through the graph from left to right.
In the figure, a single use of the matrix $\mvec{F}$ is highlighted in red. The structure of a polar code is composed of a recursive application of the building block $\mvec{F}$, arranged in $n$ layers marked by the color of the dashed rectangles. The edges on any layer in the structure are enumerated as $i{=}0,1,{\dots},N{-}1$ from top to bottom. The layers are labeled $d=1,\cdots,n$.

\subsection{Successive Cancellation (List)  Decoding}
The \gls{sc}\cite{arikan_channel_2009} decoder  estimates $\mvec{u}$ bit-by-bit in a sequential manner. In the Fig. \ref{fig:pc8} representation, the \glspl{llr} propagate from right to left and produce the decision level \gls{llr} $L_{u_i}$ which is used to estimate $\hat{u}_i$ for $i \in \mset{A}$ as:
\begin{equation}\label{eq:SC_decoding_rule}
    \hat{u}_i= h(L_{u_i}) =
        \begin{cases}
        0	&	L_{u_i} \geq 0 	\\
        1	&	\text{otherwise},
        \end{cases}		
\end{equation}
where $h(a)=\frac{1}{2}(1-\sgn{a)}$ denotes hard decision decoding with $\sgn{a}$ producing $-1$ when $a<0$ and $+1$ otherwise. 
For $i {\notin} \mset{A}$, the frozen bit value is known to the decoder.

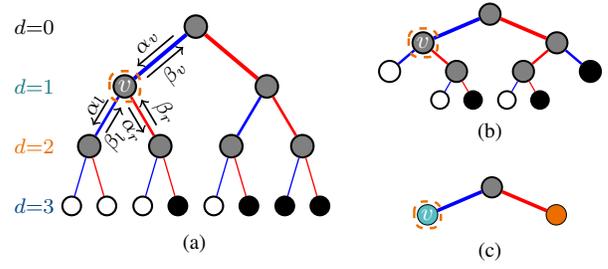
\begin{figure}[t]
  \begin{minipage}{0.65\columnwidth}
    \centering
    \subfloat[]{\label{fig:sc-tree}\hspace{-20pt}\resizebox{0.75\columnwidth}{!}{\rotatebox{90}{\begin{tikzpicture}[baseline = (0_7.center),
        level/.style={level distance = 8mm},
        level 1/.style={sibling distance=19mm, edge from parent/.style={draw,blue,line width=1.5pt}},
        level 2/.style={sibling distance=9.5mm, edge from parent/.style={draw,blue,line width=1pt}},
        level 3/.style={sibling distance=4.7mm, edge from parent/.style={draw,blue,line width=0.5pt}},
        ]

\tikzset{
frozen/.style={thick,draw=black,fill=white,minimum size=2.5mm,circle, inner sep=0},
fullspace/.style={thick,draw=black,fill=black,minimum size=2.5mm,circle, inner sep = 0},
mixed/.style={thick,draw=black,fill=gray,minimum size=3.0mm,circle, inner sep = 0},
phantom/.style={draw=white,fill=white,minimum size=2.5mm,circle, inner sep = 0},
}

\tikzset{
parallel segment/.style={
   segment distance/.store in=\segDistance,
   segment pos/.store in=\segPos,
   segment length/.store in=\segLength,
   to path={
   ($(\tikztostart)!\segPos!(\tikztotarget)!\segLength/2!(\tikztostart)!\segDistance!90:(\tikztotarget)$) -- 
   ($(\tikztostart)!\segPos!(\tikztotarget)!\segLength/2!(\tikztotarget)!\segDistance!-90:(\tikztostart)$)  \tikztonodes
   }, 
   segment pos=.5,
   segment length=2.5ex,
   segment distance=-1mm,
},
}

\tikzset{green dotted/.style={draw=green!50!black, line width=0.75pt,
    dash pattern=on 3pt off 3pt,
    inner sep=0.4mm, rectangle, rounded corners}};

\node[mixed] (p){} [grow=left]
	child {node[mixed] (v){\rotatebox{-90}{\textcolor{white}{$v$}}}	
		child {node[mixed] (1_0){}
			child {node[frozen] (a0_0){}
			}
			child {node[frozen] (a0_1){} edge from parent[red]
			}
		}
		child {node[mixed] (1_2){} edge from parent[red]
			child {node[frozen] (0_2){}
			}
			child {node[fullspace] (0_3){} edge from parent[red]
			}
		}
	}
	child {node[mixed] (2_1){} edge from parent[red]	
		child {node[mixed] (cl){}
			child {node[frozen] (0_4){}
			}
			child {node[fullspace] (0_5){} edge from parent[red]
			}
		}
		child {node[mixed] (cr){} edge from parent[red]
			child {node[fullspace] (0_6){}
			}
			child {node[fullspace] (0_7){} edge from parent[red]
			}
		}
	}
;

\draw[->,line width=0.65pt] (p) to[parallel segment,segment length=4ex] node[above right=-02.0mm] {\rotatebox{-40}{\footnotesize $\alpha_v$}} (v);
\draw[->,line width=0.65pt] (v) to[parallel segment,segment length=4ex] node[below left=-2.0mm] {\rotatebox{-40}{\footnotesize $\beta_v$}} (p);

\draw[->,line width=0.65pt] (1_0) to[parallel segment] (v) {};
\node at ($(1_0.south)-(-0.08,0.17)$) {\rotatebox{-30}{\footnotesize $\beta_l$}};
\draw[->,line width=0.65pt] (v) to[parallel segment] node[above right=-2.0mm] {\rotatebox{-20}{\footnotesize $\alpha_l$}} (1_0);

\draw[->,line width=0.65pt] (v) to[parallel segment] (1_2) {};
\node at ($(1_2.north)+(+0.2,0.23)$) {\rotatebox{215}{\footnotesize $\alpha_r$}};
\draw[->,line width=0.65pt] (1_2) to[parallel segment] node[below =-1.0mm] {\rotatebox{215}{\footnotesize $\beta_r$}} (v);

\tikzset{Nv_4/.style={draw=orange, line width=1pt,
    dash pattern=on 3pt off 3pt,
    inner sep=0.4mm, rectangle, rounded corners}};
\node (g_concat) [Nv_4, fit = (v)] {};

\node (d3) at ($(a0_0)+(-0.0,0.50)$) {\rotatebox{-90}{\color{darkblue}\footnotesize $d{=}3$}};
\node (d2) at ($(d3)+(8mm,0)$) {\rotatebox{-90}{\color{orange}\footnotesize $d{=}2$}};
\node (d1) at ($(d2)+(8mm,0)$) {\rotatebox{-90}{\color{darkgreen}\footnotesize $d{=}1$}};
\node (d0) at ($(d1)+(8mm,0)$) {\rotatebox{-90}{\footnotesize $d{=}0$}};
\end{tikzpicture}}}}
  \end{minipage}
  \hspace{-30pt}
  \begin{minipage}{0.42\columnwidth}
    \centering
    \subfloat[]{\label{fig:ssc-tree}\resizebox{0.8\columnwidth}{!}{\rotatebox{90}{\begin{tikzpicture}[baseline = (v.center),
        level/.style={level distance = 4.0625mm},
        level 1/.style={sibling distance=19mm, edge from parent/.style={draw,blue,line width=1.5pt}},
        level 2/.style={sibling distance=9.5mm, edge from parent/.style={draw,blue,line width=1pt}},
        level 3/.style={sibling distance=4.7mm, edge from parent/.style={draw,blue,line width=0.5pt}},
        ]

\tikzset{
frozen/.style={thick,draw=black,fill=white,minimum size=3mm,circle, inner sep=0},
fullspace/.style={thick,draw=black,fill=black,minimum size=3mm,circle, inner sep = 0},
mixed/.style={thick,draw=black,fill=gray,minimum size=3mm,circle, inner sep = 0},
phantom/.style={draw=white,fill=white,minimum size=3mm,circle, inner sep = 0},
frozen_leaf/.style={thick,draw=black,fill=white,minimum size=2.5mm,circle, inner sep=0},
fullspace_leaf/.style={thick,draw=black,fill=black,minimum size=2.5mm,circle, inner sep = 0},
}

\tikzset{green dotted/.style={draw=green!50!black, line width=1pt,
    dash pattern=on 3pt off 3pt,
    inner sep=0.4mm, rectangle, rounded corners}};

\node[mixed] (p){} [grow=left]
	child {node[mixed] (v){\rotatebox{-90}{\textcolor{white}{$v$}}}	
		child {node[frozen] (1_0){}
		}
		child {node[mixed] (1_2){} edge from parent[red]
			child {node[frozen_leaf] (0_2){}
			}
			child {node[fullspace_leaf] (0_3){} edge from parent[red]
			}
		}
	}
	child {node[mixed] (2_1){} edge from parent[red]	
        child {node[mixed] (1_3){} edge from parent[red]
			child {node[frozen_leaf] (0_4){}
			}
			child {node[fullspace_leaf] (0_5){} edge from parent[red]
			}
		}
      child {node[fullspace] (1_3){}
    		}
	}
;

\tikzset{Nv_4/.style={draw=orange, line width=1pt,
    dash pattern=on 3pt off 3pt,
    inner sep=0.4mm, rectangle, rounded corners}};
\node (g_concat) [Nv_4, fit = (v)] {};

\end{tikzpicture}}}}
    \vspace{-1pt}
    \subfloat[]{\makebox[\columnwidth][c]{\label{fig:fastssc-tree}\resizebox{0.55\columnwidth}{!}{\rotatebox{90}{\begin{tikzpicture}[baseline = (v.center),
        level/.style={level distance = 4.0625mm},
        level 1/.style={sibling distance=19mm, edge from parent/.style={draw,blue,line width=1.5pt}},
        level 2/.style={sibling distance=9.5mm, edge from parent/.style={draw,blue,line width=1pt}},
        level 3/.style={sibling distance=4.7mm, edge from parent/.style={draw,blue,line width=0.5pt}},
        ]

\tikzset{
    frozen/.style={thick,draw=black,fill=white,minimum size=3mm,circle, inner sep=0},
    fullspace/.style={thick,draw=black,fill=black,minimum size=3mm,circle, inner sep = 0},
    mixed/.style={thick,draw=black,fill=gray,minimum size=3mm,circle, inner sep = 0},
    phantom/.style={draw=white,fill=white,minimum size=3mm,circle, inner sep = 0},
    R0/.style={thin,draw=black,fill=white,minimum size=3mm,circle, inner sep = 0},
	R1/.style={thin,draw=black,fill=black,minimum size=3mm,circle, inner sep = 0},
	Rep/.style={thin,draw=black,fill=lightgreen,minimum size=3mm,circle, inner sep = 0},
	SPC/.style={thin,draw=black,fill=orange,minimum size=3mm,circle, inner sep = 0}
}

\tikzset{green dotted/.style={draw=green!50!black, line width=1pt,
    dash pattern=on 3pt off 3pt,
    inner sep=0.4mm, rectangle, rounded corners}};

\node[mixed] (p){} [grow=left]
	child {node[Rep] (v){\rotatebox{-90}{\textcolor{white}{$v$}}}	
	}
	child {node[SPC] (2_1){} edge from parent[red]	
	}
;

\tikzset{Nv_4/.style={draw=orange, line width=1pt,
    dash pattern=on 3pt off 3pt,
    inner sep=0.4mm, rectangle, rounded corners}};
\node (g_concat) [Nv_4, fit = (v)] {};

\end{tikzpicture}}}}}
  \end{minipage}
  \caption[Decoder-tree representations of a polar code with $N=8$ and $R=1/2$ with $\mathcal{A}{=} \{ 3,5,6,7\}$. (a) \gls{sc}, (b) Simplified \gls{sc} (c) Fast simplified \gls{sc}.]{Decoder-tree representations of a polar code with $N=8$ and $R=1/2$ with $\mathcal{A}{=} \{ 3,5,6,7\}$. (a) \gls{sc}, (b) Simplified \gls{sc} (c) Fast simplified \gls{sc}. Node types: \tikz{\draw[fill=black!50,line width=0.5pt,anchor=base]  circle(2.5pt);} $SC$, \tikz{\draw[fill=white,line width=0.5pt,anchor=base]  circle(2.5pt);} $R0$,  \tikz{\draw[fill=black,line width=0.5pt,anchor=base]  circle(2.5pt);} $R1$, \tikz{\draw[fill=lightgreen,line width=0.5pt,anchor=base]  circle(2.5pt);} $Rep$, \tikz{\draw[fill=orange,line width=0.5pt,anchor=base]  circle(2.5pt);} $SPC$.}
  \label{fig:pc_8_4}
  \vspace{-10pt}
\end{figure}

The \gls{sc} decoding can also be represented by a binary decoding tree\cite{Alamdar_SSC_2011} like Fig. \ref{fig:sc-tree}. In this representation, all the decoding operations that can be performed in parallel, i.e., the dashed rectangles in Fig. \ref{fig:pc8}, are condensed into a single decoder node. 
The leaf nodes correspond to the encoder input $\mvec{u}$ with frozen and information bits denoted by white and black color, respectively.
Moreover, the layer label $d$ can be interpreted as depth in the tree.

The decoding schedule activates the decoder nodes in top to bottom and left to right order. 
Upon activation, an $SC$ node $v$ receives $N_v$ \glspl{llr} $\mvecm{\alpha}_v{=}[\alpha_{v,0},\alpha_{v,1},{\cdots},\alpha_{v,N_v{-}1}]^T$ from its parent node and is responsible for providing $\mvecm{\beta}_v{=}[\beta_{v,0},\beta_{v,1},\cdots,\beta_{v,N_v-1}]^T$, a bit-valued vector, to its parent node. 
It computes $\sfrac{N_v}{2}$ \glspl{llr}  for its left child as\cite{Leroux2011_LLR_SC}:
\begin{equation}\label{eq:sc:f_boxplus}
    \alpha_{l,i'} = \alpha_{v,i'} \boxplus \alpha_{v,i' + \sfrac{N_v}{2}},
\end{equation}
with $0{\leq}i'{<}\sfrac{N_v}{2}$, and activates its left child. 
The boxplus operation of \eqref{eq:sc:f_boxplus} can be approximated as:
\begin{equation}\label{eq:sc:f_minsum}
    \alpha_{l,i'} = \sgn{\alpha_{v,i'}\alpha_{v,i' + \sfrac{N_v}{2}}} \min(|\alpha_{v,i'}|, |\alpha_{v,i' + \sfrac{N_v}{2}}|).
\end{equation}
The node then waits for the left child to produce its decoding output  $\mvecm{\beta}_l$.
Once $\mvecm{\beta}_l$ is available, the node $v$ activates its 
right child by sending it the \gls{llr} vector  $\mvecm{\alpha}_r$, obtained as:
\begin{equation}\label{eq:sc:g}
    \alpha_{r,i'}=(-1)^{\beta_{l,i'}} \alpha_{v,i'} + \alpha_{v,i' + \sfrac{N_v}{2}}.
\end{equation} 
With the output of the right child $\mvecm{\beta}_r$ at hand,
node $v$ computes its decoding output $\mvecm{\beta}_v$ as:
\begin{equation}\label{eq:sc:node_output}
    \beta_{v,i} =
        \begin{cases}
        \beta_{l,i}\oplus\beta_{r,i} 	&	\text{if } i<\sfrac{N_v}{2}\\
        \beta_{r,i-\sfrac{N_v}{2}}	&	\text{otherwise}.
        \end{cases}		
\end{equation}
If a leaf node is activated, $\mvecm{\alpha}_v{=}L_{u_i}$ and $\mvecm{\beta}_j{=}\hat{u}_i$ with the help of \eqref{eq:SC_decoding_rule}. 
For the root node, $\mvecm{\alpha}_v$ are the channel \glspl{llr} while $\mvecm{\beta}_v{=}\hat{\mvec{x}}$, i.e., an estimate  for the transmitted codeword. 
For a systematic polar code,  the information bits are directly retrieved from $\hat{\mvec{x}}$. In the non-systematic setting of Fig.~\ref{fig:pc8}, the estimated encoder input can be obtained as $\hat{\mvec{u}} {=} \mvec{F}^{\otimes n} \hat{\mvec{x}}$.

Contrary to the \gls{sc} decoder, an \gls{scl}\cite{tal_list_2015} decoder keeps track of multiple candidate outputs. Every time a leaf node with $i \in \mset{A}$ is encountered, the list decoder pursues both estimates of $\hat{u}_i{=}0$ and $\hat{u}_i{=}1$, doubling the number of candidates. 
At the $i$th leaf node, the $j$th candidate in the list is assigned a path penalty metric\cite{balatsoukas-stimming_llr-based_2015} according to $\mu_{i,j}=\mu_{i-1,j}+\Delta\mu_{i,j}$ with
\begin{equation}\label{eq:scl:pm_exact}
    \Delta\mu_{i,j}= \log (1+e^{ -(1-2\hat{u}_{i,j}) L_{u_{i,j}} } ),  
\end{equation}
for $j {\in }\left\lbrace 0,1\dots N_L{-}1 \right\rbrace$ and $\mu_{-1,j}{=}0$. 
If the number of candidates in the list exceeds the specified maximum list size $N_L$, only the $N_L$ most likely candidates are retained and the rest are dropped from the list.
The $\Delta\mu_{i,j}$ is approximated as \cite{balatsoukas-stimming_llr-based_2015}:
\begin{equation}\label{eq:scl:pm_approx}			
	\Delta \mu_{i,j}= 
 (\hat{u}_{i,j}\oplus h(L_{u_{i,j}}))|L_{u_{i,j}}|
\end{equation}
It can be seen that  \eqref{eq:scl:pm_approx} penalizes a path where the estimate $\hat{u}_{i,j}$ does not match the hard decision on the LLR $L_{u_i}$.
Once the $N_L$ candidates for $\mvec{u}$ are produced by the root node, the path with smallest metric is selected as the decoder output.

\subsection{Fast Decoding Nodes}\label{sec:fssc_nodes}
The fast polar decoders identify and exploit special nodes in the decoder tree of a polar code
These are in fact small polar codes of size $N_v{<}N$. 
These nodes can produce their output $\mvecm{\beta}_v$ directly from their input $\mvecm{\alpha}_v$ without traversing the decoding tree down to leaf nodes. 
The number and position of frozen bits in the $N_v$ leaf nodes rooted in a special node defines it type.
This work considers the following special nodes:
\begin{itemize}
    \item \gls{noder0} Node: When all the $N_v$ leaf nodes rooted in a decoder node $v$ are frozen, it is classified as a rate 0 node\cite{Alamdar_SSC_2011}.
    \item $R1$ Node: A decoder node is classified as a rate 1 node when all of its associated leaf nodes correspond to information bits\cite{Alamdar_SSC_2011}. 
    \item $Rep$ Node: When all except the right-most (i.e., last) leaf nodes of $v$ are frozen, it is classified as a rate $\sfrac{1}{N_v}$ repetition node\cite{Sarkis_FastSSC_2014}.
    \item $SPC$ Node: A node $v$ with only left-most (i.e., first) leaf node being frozen is identified as a rate $\sfrac{(N_v-1)}{N_v}$ \glsentrylong{spc} node\cite{Sarkis_FastSSC_2014}.
\end{itemize}

Figs.~\ref{fig:ssc-tree} and \ref{fig:fastssc-tree} show the \gls{ssc} and \gls{fssc} decoding trees for the rate $0.5$ polar code of Fig.~\ref{fig:sc-tree}, respectively. 
An \gls{scl} decoder that uses the \gls{fssc} decoding tree is said to use  \gls{fssc} schedule and, henceforth, referred to as a \glsentrylong{fssc} list (FSCL) decoder.

\subsection{Fast Successive Cancellation List Decoding}\label{subsec:fssc_dec}
This section briefly revisits the decoding procedure of the special nodes under the \gls{scl} decoding \cite{Sarkis_2016_FSSCL,Hashemi2017_FSSL_list_size}. Except for the node size $N_v$, the subscript $v$ is dropped for the sake of brevity. 
Assume there are $N_L$ paths in the list when a decoder node is activated. 
Each path enters the node carrying a path metric $\mu_j$ and an \gls{llr} vector $\mvecm{\alpha}_j$ with $0{\leq}j{<}N_L$.
Further, $\Delta \mu_{j} {=} \sum_{i=0}^{N_v{-}1}|\beta_{i,j} {-} h(\alpha_{i,j})| |\alpha_{i,j}|$\cite{Sarkis_2016_FSSCL} denotes the path metric update for the decoding candidate $\mvecm{\beta}_j{=}[\beta_{0,j},\beta_{1,j},{\cdots},\beta_{N_v{-}1,j}]^T$ of a special node. 
The node has to produce a maximum of $N_L$ outputs $\mvecm{\beta}_j$ with smallest $\mu_j+\Delta\mu_j$. Further, the hardware-friendly approximate formulation for the path metric update is used in the following.

Each outputs $\mvecm{\beta}_j$ of an \gls{noder0} node is an all-zero vector. 
Under list decoding, the node does not increase the number of paths in the list but the path metric of each  path is updated by
\begin{equation}\label{eq:fssc:pm_r0}
    \Delta \mu_j = \sum_{i=0}^{N_v-1}h(\alpha_{i,j}) |\alpha_{i,j}|.
\end{equation}

For a \gls{noderep} node, valid values of $\mvecm{\beta}_j$ are either all-zero or all-one vector of length $N_v$. Therefore, each path entering a \gls{noderep} node produces two forks, with metric updates:
\begin{equation}\label{eq:fssc:pm_rep}
    \Delta \mu_j =
        \begin{cases}
        \sum_{i=0}^{N_v-1}h(\alpha_{i,j}) |\alpha_{i,j}| 	&	\forall \beta_j =0\\
        \sum_{i=0}^{N_v-1}|1-h(\alpha_{i,j})| |\alpha_{i,j}|	&	\forall \beta_j =1.
        \end{cases}		
\end{equation}

The \gls{noder1} and \gls{nodespc} nodes have $2^{N_v}$ and $2^{N_v-1}$ possible outputs, respectively. 
Both nodes obtain the path increment $\Delta \mu_{j}$ in a limited number of steps by using the \gls{ml} candidate for each paths entering the node, computed as
\begin{equation}\label{eq:fssc:beta_ML}
    \mvecm{\beta}_j^{ML} = h(\mvecm{\alpha}_{j}).
\end{equation}
Further, let $\mvecm{\alpha}'_j{=}[\alpha'_{0,j},\cdots,\alpha'_{i',j},{\cdots},\alpha'_{N_v-1,j}]^T$ represent the \glspl{llr} $\mvecm{\alpha}_j$  sorted w.r.t. reliability, i.e., $|\alpha_{i,j}|$. The $N_L$ promising candidates to be  retained in the decoding list are efficiently generated as follows\cite{Hashemi2017_FSSL_list_size}:

An \gls{noder1} node starts with initializing $\Delta \mu_j {=}0$ for each \gls{ml} candidate entering the node. The \gls{noder1} decoder sifts concurrently through $\mvecm{\alpha}'_j$ in the order $0\cdots i'_j{\cdots}s_{\text{R1}}$ with $s_{\text{R1}}{=}\min\left( N_L-1,N_v\right)$. At each position $i'_j$, the decoder splits the corresponding \gls{ml} path $\mvecm{\beta}^{ML}_j$ and increments the metric update of the new fork by $|\alpha_{i'_j}|$. Once the least reliable $s_{\text{R1}}$  positions in the \gls{ml} candidates are considered, no further path splitting is done and the \gls{noder1} decoder retains the $N_L$ decoding paths with smallest path metrics $\mu_j+\Delta\mu_j$ in the list.

An \gls{nodespc} decoder considers $s_{\text{SPC}}{=}\min\left( N_L,N_v\right)$\cite{Hashemi2017_FSSL_list_size} least reliable positions in $\mvecm{\alpha}_j$ for path splitting. Let the least reliable \gls{llr} position, i.e., $i'_j=0$ in $\mvecm{\alpha}'_j$, be denoted by $i_{j,\min}$. 
First, the parity of $j$th path is computed as
\begin{equation}
    \gamma_j = \bigoplus_{i=0}^{N_v-1} \mvecm{\beta}^{ML}_j,
\end{equation}
and its metric update  is initialized to $\gamma_j |\alpha_{i_{j,\min}}|$.
The \gls{nodespc} decoder then goes through the remaining $s_{\text{SPC}}-1$ least reliable positions in the order $1\cdots i'_j{\cdots}s_{\text{SPC}}$.  At each step $i'_j$, a fork is created from the surviving \gls{ml} paths in the list with a metric update incremented by $|\alpha_{i'_j}| + (1-2\gamma_j) |\alpha_{i_{j,\min}}|$. Once the $N_L$ surviving paths in the list are at hand after processing the $s_{\text{SPC}}$ positions, the least reliable bit in each surviving path $\mvecm{\beta}_j$ is set to preserve the even-parity as
\begin{equation}
	\beta_{i_{j,\min}}=\bigoplus_{i\in \{0,\cdots,N_v-1 \} \setminus i_{j,\min}}\beta_{i,j}.
\end{equation}

The \gls{noder0}, \gls{noder1} and \gls{noderep} nodes in an \gls{fssc}L decoder preserve the error rate of the conventional \gls{scl} decoding\cite{Sarkis_2016_FSSCL,Hashemi2016_SSCL_SCL_equivalence,Hashemi2017_FSSL_list_size}. However, the \gls{nodespc} decoder leads to a minute degradation except for list size $N_L{=}2$\cite{Hashemi2017_FSSL_list_size}.

\section{Finite Alphabet Decoders}\label{sec:fa_decoders}
Finite alphabet decoders are a family of quantized decoders that replace \glspl{llr} with $w$-bit integer-valued messages, say $t$, in order to achieve a reduced space complexity. 
Each  message $t$ belongs to a finite alphabet $\mset{T}$ of size $|\mset{T}|{=}2^w$ and embeds reliability information w.r.t. a certain bit $x$. In other words, the message $t$ corresponds to an \gls{llr} $L_x(t)$.
In this work, the \gls{ib}\cite{tishby2000information} method is used to design  finite alphabet polar decoders.%

The \gls{ib} framework compresses an observation $Y{=}y{\in}\mset{Y}$ into a compact form $T{=}t{\in}\mset{T}$ by designating a quantity of primary relevance $X{=}x$. 
For the decoder design, $x$ is some bit value, i.e., $x{\in}\{0,1\}$.
The framework offers algorithms which accept the joint distribution $p(x,y)$ and produce a deterministic compression mapping $p(t|y)$. 
The key idea  is to determine a mapping $p(t|y)$ which maximizes the \textit{relevant} mutual information $I(X;T)$ with the constraint $|\mset{T}|{<}|\mset{Y}|$. 
The deterministic mapping $p(t|y)$ represents the compression operation in the form of an \gls{lut}.
An \gls{ib} algorithm also provides the distribution $p(x|t)$ which is used to compute the \gls{llr} $L_x(t)=\log \frac{p(x=0|t)}{p(x=1|t)}$ associated with each $t\in\mset{T}$.%

For designing finite alphabet decoders, an integer valued  alphabet  $\mset{T}{=}\{0,1,\dots,|\mathcal{T}|{-}1\}$ is used here.
The alphabet  $\mathcal{T}$ is chosen such that it is  sorted w.r.t. the underlying LLRs, i.e., $L_x(t{=}0){<}L_x(t{=}1){<}\ldots{<}L_x(t{=}|\mset{T}|{-}1)$. Moreover, the LLRs $L_x(t)$ are forced to exhibit odd symmetry such that
\begin{align*}
	L_{x}(t = \sfrac{|\mathcal{T}|}{2} - 1 - k)= - L_{x}(t = \sfrac{|\mathcal{T}|}{2} + k),
	\label{eq:quanti_symmetry}
\end{align*}
where $ k = 0, \cdots , \sfrac{|\mathcal{T}|}{2} - 1$. 

In the following, the design  of quatized polar decoder  using the \gls{ib} method from \cite{shah_design_2019,shah_coarsely_2019} is revisited briefly.
\subsection{Decoder Design}\label{subsec:ib_decoder_design}

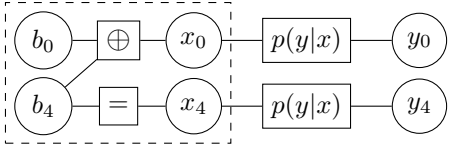
\begin{figure}[tbp]
	\centering      
	\begin{tikzpicture}[yscale=0.35, xscale=1, node distance=0.3cm, auto]
	
		\def \nodesize {0.5} %
		\def \VertDist {2.5}
		\def \HorDist {1}
		\def \HorDisplace {0.5}
		
		\node (U0) at (0-\HorDist,0) [draw, circle, minimum width = \nodesize cm]{$b_0$};
		\node (U1) at (0-\HorDist,0-\VertDist) [draw, circle, minimum width = \nodesize cm]{$b_4$};%
		\node (Cn0) at (0,0) [draw, rectangle, minimum width = \nodesize cm,minimum height = \nodesize cm]{\large$\oplus$};
		\node (Cn1) at (0,0-\VertDist) [draw, rectangle, minimum width = \nodesize cm,minimum height = \nodesize cm]{$=$};	
		\node (X0) at (\HorDist,0) [draw, circle, minimum width = \nodesize cm]{$x_0$};
		\node (X1) at (\HorDist,0-\VertDist) [draw, circle, minimum width = \nodesize cm]{$x_4$};
		\node (Cn2) at (2*\HorDist +\HorDisplace,0) [draw, rectangle, minimum width = \nodesize cm,minimum height = \nodesize cm]{$p(y|x)$};
		\node (Cn3) at (2*\HorDist +\HorDisplace,0-\VertDist) [draw, rectangle, minimum width = \nodesize cm,minimum height = \nodesize cm]{$p(y|x)$};	
		\node (Y0) at (3*\HorDist +2*\HorDisplace,0) [draw, circle, minimum width = \nodesize cm]{$y_0$};
		\node (Y1) at (3*\HorDist +2*\HorDisplace,0-\VertDist) [draw, circle, minimum width = \nodesize cm]{$y_4$};

		\draw[-] (U0) -- (Cn0);
		\draw[-] (U1) -- (Cn1);
		\draw[-] (Cn0) -- (X0);
		\draw[-] (Cn1) -- (X1);
		\draw[-] (U1) -- (Cn0);
		\draw[-] (X0) -- (Cn2) --(Y0);
		\draw[-] (X1) -- (Cn3) --(Y1);
		
		\node[draw,dashed,fit=(U0) (X1)] {};
	
	\end{tikzpicture}	
    \caption{ Factor graph of the building block and the transmission channel.}
    \label{fig:Tannergraph2.1}
   \vskip -10pt
\end{figure}

The process of generating \glspl{lut} for decoding is explained using Fig.\,\ref{fig:Tannergraph2.1} on a single building block (highlighted in red color in Fig.~\ref{fig:pc8}). 
In Fig.\,\ref{fig:Tannergraph2.1}, $p(y|x)$ represents  a quantized binary input AWGN channel. 

First, the \gls{ib} method is used to quantize the underlying AWGN channel such that $y_i{\in} \mathcal{T}$\cite{lewandowsky_information-optimum_2018}. The channel quantizer provides the \glspl{llr} $L_x(y_i)$. With the quantized channel outputs $\mvec{y}{=}[y_0, y_4]^T$ and $L_x(y_i)$ at hand, the \gls{ib} algorithm constructs the observed \gls{llr} space using \eqref{eq:sc:f_boxplus} for the upper branch in Fig.~\ref{fig:Tannergraph2.1}. The algorithm places $|\mset{T}|-1$ boundaries in the sorted observed \gls{llr} space and optimizes them such that $I(B_0;T_0)$ is maximized. This results in a compression mapping  $p(t_0|\mvec{y})$ with $t_0{\in}\mset{T}$, as well as the \glspl{llr} $L_{b_0}(t_0)$. 
The \gls{lut} $p(t_0|\mvec{y})$  compresses the input alphabet of size $2^{2w}$ to an output alphabet of size $|\mset{T}|{=}2^w$. 

For the lower branch in  Fig.~\ref{fig:Tannergraph2.1}, the observed \gls{llr} space is constructed from the \glspl{llr} $L_x(y_i)$ of the quantized channel outputs $\mvec{y}{=}[y_0, y_4]^T$, and $\hat{b}_0\in\{0,1\}$ according to \eqref{eq:sc:g}. 
The $|\mset{T}|-1$ boundaries in the sorted observed \gls{llr} space are optimized such that $I(B_4;T_4)$ is maximized. 
The result is an \gls{lut}  $p(t_4|\mvec{y},\hat{b}_0)$ with $t_4{\in}\mathcal{T}$ which compresses the input alphabet of size $2^{2w+1}$ to an output alphabet of size $2^w$, as well the \glspl{llr} $L_{b_4}(t_4)$.

The \glspl{lut} $p(t_0|\mvec{y})$ and $p(t_4|\mvec{y},\hat{b}_0)$ are valid for the upper and the lower branch updates, respectively, of each building block on the first layer (blue dashed rectangle) in Fig. \ref{fig:pc8}. 
The aforementioned procedure is recursively extended to the next layers in Fig.\ref{fig:pc8} to obtain compression mapping for each building block branch as detailed in \cite{shah_design_2019,shah_coarsely_2019}.  As a result, a total of $2N{-}2$, same as the number of edges in Fig.\,\ref{fig:sc-tree}, distinct compression mappings are obtained for a polar code of length $N$. 
The AWGN channel quantizer as well as the decoder \glspl{lut} are designed offline for a certain  $E_b/N_0$, which is referred to as the design  $E_b/N_0$ of the finite alphabet decoder designed using the \gls{ib} method.

\subsection{LUT based Successive Cancellation List Decoding}\label{subsec:ib_scl_decodeing}
 The decoder tree representation of Fig.~\ref{fig:sc-tree} is used to explain the \gls{lut}-based finite alphabet decoding. Half of the $2N{-}2$ mappings utilize \eqref{eq:sc:f_boxplus} in their design and correspond to the left edges in Fig. \ref{fig:sc-tree}. The remaining $N{-}1$ \glspl{lut} use \eqref{eq:sc:g} during their design and correspond to the right edges in the figure.

At the start of the finite alphabet decoding, the root node receives the  integer-valued quantized channel outputs $\mvecm{\alpha}_v {=} [y_0,\dots,y_7]^T$. 
The root node determines $\mvecm{\alpha}_l$ for its left child using the \gls{lut} $p(t_{i'}|y_{i'},y_{i'+\sfrac{N_v}{2}})$ with $i'{=}0,{\dots},\sfrac{N_v}{2}$, i.e., the mapping $p(t_0|\mvec{y})$ with  adjusted labels. 
Since the  \gls{lut} $p(t_0|\mvec{y})$ replaces the \gls{llr} arithmetic of \eqref{eq:sc:f_boxplus}, it is referred to  as a \textit{decoding} table. The list of \glspl{llr} $L_{b_0}(t_0)$ is referred to as the \textit{translation} table for $t_{i'}$ as it translates each integer-valued message into its reliability information.

When the left child produces its decoding output $\mvecm{\beta}_l$, the root node computes $\mvecm{\alpha}_r$ for its right child using the \gls{lut} $p(t_4|\mvec{y},\hat{b}_0)$ as  $p(t_{i'}|y_{i'},y_{i'{+} \sfrac{N_v}{2}},\beta_{l,i'})$ with $i'{=}0,{\dots},\sfrac{N_v}{2}$. As the \gls{lut} replaces the LLR arithmetic \eqref{eq:sc:g}, it is termed a decoding table and it comes with its own translation table, i.e., the list of \glspl{llr} $L_{b_4}(t_4)$. After the right child produces its decoding output $\mvecm{\beta}_r$, it is combined with $\mvecm{\beta}_l$ according to \eqref{eq:sc:node_output}. 

Each node in the decoding tree uses a separate decoding table to produce integer-valued inputs for its left or right child indicated by blue or red edges, respectively, in Fig.~\ref{fig:sc-tree}. Every leaf node is reached after using a different sequence $n$ decoding tables (the sequence of $n$ edges connecting the root node to the leaf node). At each leaf node, a separate translation table is used to convert the integer messages into \glspl{llr} for path metric update according to \eqref{eq:scl:pm_exact} or \eqref{eq:scl:pm_approx}. 
Thus, although the decoder design procedure creates a translation table associated with each of the $2N-2$ decoding tables, only $N$ translation tables are used in the \gls{scl} decoding.

\section{Finite Alphabet Fast Successive Cancellation List Decoding}

This section explains how the \glspl{lut} generated for \gls{sc} schedule in Sec. \ref{sec:fa_decoders} can be used in the fast decoding schedules. The main difference arises from the use of decoding trees of Figs. \ref{fig:ssc-tree} or \ref{fig:fastssc-tree} instead of Fig. \ref{fig:sc-tree}.

The fast finite alphabet decoding starts with the root node receiving the integer-valued quantized channel outputs $y_0,\dots,y_7$. The root node activates its children at depth $d=1$ exactly as explained in Sec. \ref{subsec:ib_scl_decodeing}. Consider the instance when the left child marked $v$ at $d=1$ in Fig. \ref{fig:pc_8_4} is activated. The node $v$ receives $\mvecm{\alpha}_v {=}[t_0,\dots,t_3]^T$ from the root node. From $\mvecm{\alpha}_v$, the node produces $\mvecm{\alpha}_l$ for its left child using the \gls{lut} corresponding to the edge between the two nodes. Under the \gls{ssc} schedule, the left child is an \gls{noder0} node. The rate-0 child translates the integer-valued messages in $\mvecm{\alpha}_l$  to \glspl{llr} and produces its output $\mvecm{\beta}_l$ as explained in Sec. \ref{subsec:fssc_dec}. The right child of $v$ is a normal node which produces its output $\mvecm{\beta}_r$ by traversing the tree down to the maximum depth $d=3$.

On the other hand, the \gls{fssc} tree only has two leaf nodes that are children of the root node. Under the \gls{fssc} schedule, the node $v$ already is a special, i.e., \gls{noderep}, node. Thus, node $v$ translates $\mvecm{\alpha}_v$ into \glspl{llr} using the translation table $L_{b_0}(t_0)$ from Sec. \ref{subsec:ib_scl_decodeing} and produces $\mvecm{\beta}_v$ according to the repetition node decoding of Sec. \ref{subsec:fssc_dec}. Similarly, the right child is an \gls{nodespc} node. When activated, the \gls{nodespc} node translates the integer-valued messages it receive from the root node using  the translation table $L_{b_4}(t_4)$ from Sec. \ref{subsec:ib_scl_decodeing}. The \gls{nodespc} node produces its output from the translated \glspl{llr} as described in Sec. \ref{subsec:fssc_dec}.

The number of decoding tables required by a decoder schedule depends upon the number of edges in its respective decoding tree. For the example of Fig. \ref{fig:pc_8_4}, the \gls{sc} schedule requires $14$ decoding tables. On the other hand, the \gls{ssc} and \gls{fssc} require $10$ and only $2$ decoding tables, respectively.

The number of translation tables required for \gls{scl} decoding depends upon the number of leaf nodes in a schedule's decoder tree. For Fig. \ref{fig:pc_8_4}, the \gls{sc} schedule requires $N=8$ translation tables. The \gls{ssc} schedule requires $6$ translation tables. The \gls{fssc} schedule requires only $2$ translation tables.

 \section{Results}

 \begin{figure}[tb]
	\centering
	\begin{tikzpicture}
\pgfplotsset{legend style={font=\small}}
\begin{axis}[%
width=0.75*\columnwidth,
height=0.5*\columnwidth,
scale only axis,
xmin=0.0,
xmax=2.250,
ymode=log,
ymin=9e-04,
ymax=1,
yminorticks=true,
ylabel={Block Error Rate},
xlabel={$E_b/N_0$ in dB},
tick label style={font=\small},
label style={font=\small},
axis background/.style={fill=white},
xmajorgrids,
ymajorgrids,
yminorgrids,
xminorticks=true, minor x tick num=4,
legend style={at={(0.01,0.01)}, anchor=south west},
legend style={legend cell align=left, align=left, draw=white!15!black,font=\small,nodes={scale=0.9, transform shape}}
]

\addplot [color=black,solid,mark=+,line width=1.0pt]
table{%
0	0.968000000000000
0.250000000000000	0.842000000000000
0.500000000000000	0.665000000000000
0.750000000000000	0.411000000000000
1	0.181818181818182
1.25000000000000	0.0630252100840336
1.50000000000000	0.0149596090555500
1.75000000000000	0.00255882352941177
2	0.000326
};
\addlegendentry{LLR SCL}; %

\addplot [color=orange,solid,mark=+,line width=1.0pt]
table{%
	0	0.965000000000000
	0.250000000000000	0.848000000000000
	0.500000000000000	0.680000000000000
	0.750000000000000	0.418000000000000
	1	0.192121212121212
	1.25000000000000	0.0699579831932773
	1.50000000000000	0.0164555699611050
	1.75000000000000	0.00298039215686275
	2	0.000418
};
\addlegendentry{LLR FSCL};

\addplot [color=darkblue,solid,mark=o,line width=1.0pt]
table{%
0	0.968200000000000
0.250000000000000	0.899600000000000
0.500000000000000	0.736800000000000
0.750000000000000	0.508600000000000
1	0.259247275464545
1.25000000000000	0.100441540181073
1.50000000000000	0.0305889602278391
1.75000000000000	0.00675938443541108
2	0.00126095215110932
2.25000000000000	0.000173999826000174	
};
\addlegendentry{IB SCL ex.};%

\addplot [color=lightgreen,solid,mark=o,line width=1.0pt]
table{%
0	0.983800000000000
0.250000000000000	0.943000000000000
0.500000000000000	0.830200000000000
0.750000000000000	0.640800000000000
1	0.387200000000000
1.25000000000000	0.191119351216181
1.50000000000000	0.0682528968637772
1.75000000000000	0.0191820055486215
2	0.00398577099617239
2.25000000000000	0.000774049540871002	
};
\addlegendentry{IB SCL ap.};%

\addplot [color=darkred,solid,mark=o,line width=1.0pt]
table{%
	0	0.986333333333333
	0.250000000000000	0.947000000000000
	0.500000000000000	0.827500000000000
	0.750000000000000	0.630833333333333
	1	0.391333333333333
	1.25000000000000	0.180399392134823
	1.50000000000000	0.0621266641356847
	1.75000000000000	0.0182133659502590
	2	0.00366477695001664
	2.25000000000000	0.000651347556292018
};
\addlegendentry{IB FSCL};%

\addplot [color=darkblue,solid,mark=x,line width=1.0pt]
table{%
0	0.973500000000000
0.250000000000000	0.913000000000000
0.500000000000000	0.765500000000000
0.750000000000000	0.538500000000000
1	0.297993326978074
1.25000000000000	0.123574086870083
1.50000000000000	0.0359244084964504
1.75000000000000	0.00861177333004129
2	0.00143135168499323
2.25000000000000	0.000208499791500208
};
\addlegendentry{MSIB SCL ex.};%

\addplot [color=lightgreen,solid,mark=x,line width=1.0pt]
table{%
0	0.978142857142857
0.250000000000000	0.915571428571429
0.500000000000000	0.774714285714286
0.750000000000000	0.557142857142857
1	0.316654579354537
1.25000000000000	0.135491366657645
1.50000000000000	0.0421413482692481
1.75000000000000	0.00982272540199713
2	0.00185289061786607
2.25000000000000	0.000303361477160527	
};
\addlegendentry{MSIB SCL ap.};%

\addplot [color=darkred,solid,mark=x,line width=1.0pt]
table{%
	0	0.976250000000000
	0.250000000000000	0.910750000000000
	0.500000000000000	0.764750000000000
	0.750000000000000	0.545750000000000
	1	0.303877794767374
	1.25000000000000	0.132517866488081
	1.50000000000000	0.0409771282569742
	1.75000000000000	0.00935875072036332
	2	0.00186033510115690
	2.25000000000000	0.000324171525637645
};
\addlegendentry{MSIB FSCL};%

\end{axis}
\end{tikzpicture}%
	\vspace{-0.3cm}
	\caption{Quantized \gls{scl} decoding under \gls{sc} and \gls{fssc} schedules. $N{=}1024, R{=}0.5, N_{\text{CRC}}{=}16, N_L{=}32$, and BPSK over AWGN channel.}
	\label{fig:BLER_FSSCL_LUT}
	\vspace{-10pt}
\end{figure}
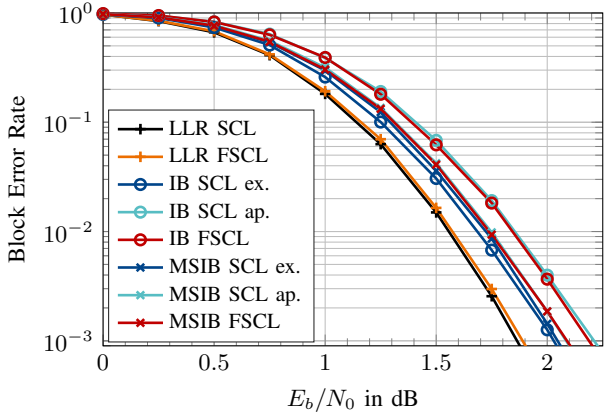

This section presents simulation results for the error correction performance of the proposed \gls{lut}-based fast \gls{scl} decoders. %
The proposed decoder is compared at at block error rate of $10^{-3}$ with  a double-precision floating-point \gls{llr}-based \gls{scl} decoder  as well as the finite alphabet decoders of \cite{shah_design_2019,shah_MSIB_2023}. 
The \gls{lut}-based decoders of \cite{shah_design_2019}, recapped in Sec.\ref{subsec:ib_decoder_design},  are labeled as \gls{ib}. The decoders from \cite{shah_MSIB_2023} which use the min-sum approximation of \eqref{eq:sc:f_minsum} during the decoder design instead of \eqref{eq:sc:f_boxplus} are labeled as MSIB.
The labels "ex." and "ap." indicate the use of exact  and approximate path metric updates according to \eqref{eq:scl:pm_exact} and \eqref{eq:scl:pm_approx}, respectively, in the \gls{sc} schedule of finite alphabet decoders. The proposed  decoders make use of the \gls{llr}-based constituent decoders for the special nodes in their \gls{fssc} schedule. 
These constituent decoders use the hardware-friendly approximate path metric updates (cf. Sec. \ref{subsec:fssc_dec}).

The \glspl{lut} for both IB and MSIB  decoders were generated with $|\mset{T}|{=}16$, i.e., 4-bit resolution.
All the simulations were performed for a codeword length of $N{=}1024$, list size of $N_L{=}32$ and CRC size of $N_{\mathrm{CRC}}{=}16$ over an AWGN channel using BPSK modulation. The code construction was adopted from 5G NR \cite{5G}.

Fig. \ref{fig:BLER_FSSCL_LUT} presents block error rates for $R{=}0.5$ where the \gls{lut}-based decoders were designed for $E_b/N_0{=}0.5$\,dB.
As reported in \cite{shah_coarsely_2019}, the use of approximate path metrics causes a visible degradation in the IB \gls{scl} decoding. Compared to the approx. $0.17$\,dB performance loss of the \gls{ib} \gls{scl} decoder, the MSIB \gls{scl} decoder suffers a smaller degradation of $0.05$\,dB due to the path metric approximation. 
Further, it can be seen that the proposed IB \gls{fssc}L and MSIB \gls{fssc}L preserve the error correction performance of IB and MSIB decoders with approximate path metrics, respectively. 

The loss due to path metric approximation can be avoided in the \gls{lut}-based decoders operating on an \gls{sc} schedule: Instead of translating the messages to \glspl{llr} and subsequently updating path metrics, the translation tables are modified such that the integer-valued messages are translated to pre-computed exact metric updates\cite{shah_coarsely_2019}. A similar workaround can be tried in the proposed fast decoders. However, we restrict ourselves in this work to employing efficient constituent decoders from existing literature that exploit the approximate path metrics. 
Within this framework, the MSIB fast SCL decoder emerges as a favorable option.
The proposed 4-bit MSIB \gls{fssc}L decoder shows a degradation of $0.2$\,dB w.r.t. the 64-bit \gls{llr}-based \gls{fssc}L decoder.
Compared to the quantized \gls{scl} decoders, the speed from using the \gls{fssc} schedule costs approx. $0.05$\,dB.

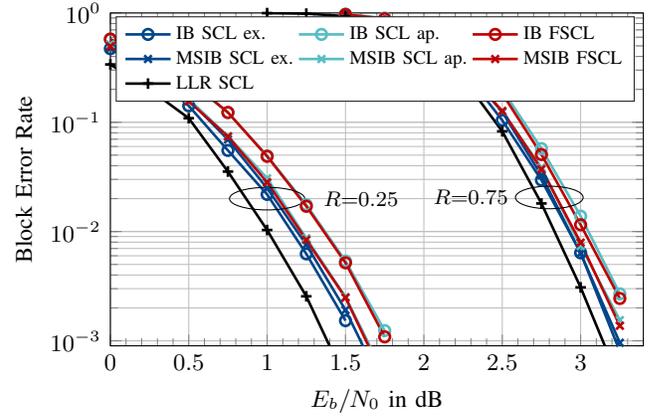
\begin{figure}[tb]
	\centering
	\begin{tikzpicture}
\begin{axis}[%
width=0.8*\columnwidth,
height=0.5*\columnwidth,
scale only axis,
xmin=0.0,
xmax=3.40,
ymode=log,
ymin=9e-04,
ymax=1,
yminorticks=true,
ylabel={Block Error Rate},
xlabel={$E_b/N_0$ in dB},
axis background/.style={fill=white},
tick label style={font=\small},
label style={font=\small},
xmajorgrids,
ymajorgrids,
yminorgrids,
xminorticks=true, minor x tick num=4,
legend columns=3,%
legend style={at={(0.01,0.99)}, anchor=north west},
legend style={legend cell align=left, align=left, draw=white!15!black,font=\small,nodes={scale=0.8, transform shape}}
]

\addplot [color=darkblue,solid,mark=o,line width=1.0pt]
table{%
0	0.471000000000000
0.250000000000000	0.288385826771654
0.500000000000000	0.141278231619414
0.750000000000000	0.0552247998357627
1	0.0218669883878751
1.25000000000000	0.00624565737610382
1.50000000000000	0.00152942335873584	
};
\addlegendentry{IB SCL ex.};%

\addplot [color=lightgreen,solid,mark=o,line width=1.0pt]
table{%
0	0.574000000000000
0.250000000000000	0.373000000000000
0.500000000000000	0.237163814180929
0.750000000000000	0.122540983606557
1	0.0492362690932727
1.25000000000000	0.0176028520828212
1.50000000000000	0.00537192818327652
1.75000000000000	0.00124000000000000
};
\addlegendentry{IB SCL ap.};%

\addplot [color=darkred,solid,mark=o,line width=1.0pt]
table{%
	0	0.577000000000000
	0.250000000000000	0.378000000000000
	0.500000000000000	0.244498777506112
	0.750000000000000	0.122950819672131
	1	0.0487487812804680
	1.25000000000000	0.0170285025336856
	1.50000000000000	0.00517193710484388
	1.75000000000000	0.00108666666666667	
};
\addlegendentry{IB FSCL};%

\addplot [color=darkblue,solid,mark=x,line width=1.0pt]
table{%
0	0.477000000000000
0.250000000000000	0.313000000000000
0.500000000000000	0.161764705882353
0.750000000000000	0.0694514343231002
1	0.0249202342301315
1.25000000000000	0.00757938448240777
1.50000000000000	0.00191401984130417
1.75000000000000	0.000351666666666667
};
\addlegendentry{MSIB SCL ex.};%

\addplot [color=lightgreen,solid,mark=x,line width=1.0pt]
table{%
0	0.515000000000000
0.250000000000000	0.348000000000000
0.500000000000000	0.162260238465526
0.750000000000000	0.0747409965466206
1	0.0307083647324793
1.25000000000000	0.00868787887253122
1.50000000000000	0.00252479263588351
1.75000000000000	0.000480000000000000	
};
\addlegendentry{MSIB SCL ap.};%

\addplot [color=darkred,solid,mark=x,line width=1.0pt]
table{%
	0	0.494000000000000
	0.250000000000000	0.324000000000000
	0.500000000000000	0.155520995334370
	0.750000000000000	0.0740009866798224
	1	0.0282592313489073
	1.25000000000000	0.00838059055228092
	1.50000000000000	0.00248340259267231
	1.75000000000000	0.000440000000000000
};
\addlegendentry{MSIB FSCL};%

\addplot [color=black,solid,mark=+,line width=1.0pt]
table{%
	0	0.338000000000000
	0.250000000000000	0.199733688415446
	0.500000000000000	0.108892921960073
	0.750000000000000	0.0353648473417423
	1	0.0103256006057686
	1.25000000000000	0.00255462642846194
	1.50000000000000	0.000440000000000000
};
\addlegendentry{LLR SCL}; %

\addplot [color=darkblue,solid,mark=o,line width=1.0pt]
table{%
	1.50000000000000	0.957000000000000
	1.75000000000000	0.829000000000000
	2	0.593000000000000
	2.25000000000000	0.302000000000000
	2.50000000000000	0.104245640636846
	2.75000000000000	0.0295504558314995
	3	0.00638328053335855
	3.25000000000000	0.000800000000000000	
};

\addplot [color=lightgreen,solid,mark=o,line width=1.0pt]
table{%
	1.50000000000000	0.975000000000000
	1.75000000000000	0.899000000000000
	2	0.705000000000000
	2.25000000000000	0.426000000000000
	2.50000000000000	0.198901769371568
	2.75000000000000	0.0574325898382523
	3	0.0138369923660904
	3.25000000000000	0.00269000000000000
};

\addplot [color=darkred,solid,mark=o,line width=1.0pt]
table{%
	1.50000000000000	0.969000000000000
	1.75000000000000	0.886000000000000
	2	0.693000000000000
	2.25000000000000	0.390000000000000
	2.50000000000000	0.183038438071995
	2.75000000000000	0.0506625177436433
	3	0.0115290796492886
	3.25000000000000	0.00244000000000000	
};

\addplot [color=darkblue,solid,mark=x,line width=1.0pt]
table{%
	1.50000000000000	0.958000000000000
	1.75000000000000	0.827000000000000
	2	0.634000000000000
	2.25000000000000	0.325000000000000
	2.50000000000000	0.124394966118103
	2.75000000000000	0.0326530612244898
	3	0.00633516923952111
	3.25000000000000	0.000960000000000000
};

\addplot [color=lightgreen,solid,mark=x,line width=1.0pt]
table{%
	2.50000000000000	0.126608551266086
	2.75000000000000	0.0384750733137830
	3	0.00736854946931804
	3.25000000000000	0.00154666666666667	
};

\addplot [color=darkred,solid,mark=x,line width=1.0pt]
table{%
	1.50000000000000	0.953000000000000
	1.75000000000000	0.852000000000000
	2	0.612000000000000
	2.25000000000000	0.315000000000000
	2.50000000000000	0.126796280642434
	2.75000000000000	0.0371919735183013
	3	0.00791908297115348
	3.25000000000000	0.00137333333333333
};

\addplot [color=black,solid,mark=+,line width=1.0pt]
table{%
	1	0.997000000000000
	1.25000000000000	0.986000000000000
	1.50000000000000	0.934000000000000
	1.75000000000000	0.760000000000000
	2	0.540000000000000
	2.25000000000000	0.252951096121417
	2.50000000000000	0.0826218672541999
	2.75000000000000	0.0180704278054794
	3	0.00308341457940852
	3.25000000000000	0.000420000000000000
};

\draw (axis cs:1,2e-2) ellipse (0.5cm and 0.15cm);
\node at (axis cs:1.6,2e-2){\footnotesize$R{=}0.25$};

\draw (axis cs:2.8,2.05e-2) ellipse (0.45cm and 0.15cm);
\node at (axis cs:2.3,2.09e-2){\footnotesize$R{=}0.75$};

\end{axis}
\end{tikzpicture}%
	\vspace{-10pt}
	\caption{Quantized \gls{scl} decoding under \gls{sc} and \gls{fssc} schedules. $N{=}1024$, $R{=}0.25$ and $0.75$, $N_{\text{CRC}}{=}16, N_L{=}32$, and BPSK over AWGN channel.}
	\label{fig:BLER_FSSCL_LUT_multiple_rates}
	\vspace{-10pt}
\end{figure}

Fig.~\ref{fig:BLER_FSSCL_LUT_multiple_rates} provides results for code rates of $0.25$ and $0.75$ with the quantized decoders designed for $0.0$\,dB and $1.75$\,dB. Trends similar to  Fig.~\ref{fig:BLER_FSSCL_LUT} can be seen in Fig.~\ref{fig:BLER_FSSCL_LUT_multiple_rates}, i.e., the proposed IB FSCL and MSIB FSCL preserving their error correction performance of \gls{sc} schedule with approximate metric updates. 
Here too, the proposed MSIB FSCL outperforms the IB FSCL.

In LLR-based FSCL decoding, the \gls{nodespc} nodes are known to cause a slight degradation \cite{Hashemi2017_FSSL_list_size}.
However,the proposed \gls{fssc}L decoders seem to either match or very slightly outperform their \gls{sc} schedule counterparts. 
Fig. \ref{fig:NodeEffect_MSIB} shows result of an experiment where the same codeword is decoded by a partially enabled fast schedule with only certain types of the special nodes retained in the decoder tree and replacing the rest with $SC$ nodes. 
The labels in \ref{fig:NodeEffect_MSIB} indicate the type of nodes enabled in the \gls{fssc} schedule.
In the case of Fig.~\ref{fig:BLER_NodeEffect_MSIB_individual}, only one type of the nodes, i.e.,  \gls{noder0}, \gls{noder1}, \gls{noderep} or \gls{nodespc}, are enabled while  multiple node types are enabled  in the case of Fig. \ref{fig:BLER_NodeEffect_MSIB}.  Fig.~\ref{fig:BLER_NodeEffect_MSIB_individual} shows that the \gls{nodespc} type nodes do cause slight degradation. The same suggested is by Fig.~\ref{fig:BLER_NodeEffect_MSIB} when \gls{noder0}, \gls{noder1} and \gls{noderep} nodes are enables but \gls{nodespc} nodes are not. 

The translation of integer-valued messages into \glspl{llr} always happens at an earlier depth $d{<}n$ in the \gls{fssc} schedule as seen in the compact representation of \gls{fssc} schedule for the rate $R{=}0.5$ polar code in Fig.~\ref{fig:FSSCL_schedule_depth}. Instead of the huge \gls{fssc} decoding tree for $N{=}1024$, Fig.~\ref{fig:FSSCL_schedule_depth} presents the schedule as a sequence of constituent decoders from $i_v{=}1$ to $i_v{=}86$ together with depth $d$ of each node in the tree.
The early translation of messages cuts back the loss caused by the \gls{nodespc} nodes as messages at depth $d{<}n$ are subject to less compression by the \gls{ib} framework.

\begin{figure}
	\begin{minipage}{0.49\columnwidth}
		\centering		
		\subfloat[]{\label{fig:BLER_NodeEffect_MSIB_individual}{\begin{tikzpicture}
\pgfplotsset{legend style={font=\small}}
\begin{axis}[%
width=0.65*\columnwidth,
height=1*\columnwidth,
scale only axis,
xmin=1.0,
xmax=2.25,
ymode=log,
ymin=8e-04,
ymax=1,
yminorticks=true,
xlabel={$E_b/N_0$ in dB},
ylabel near ticks, yticklabel pos=right,
axis background/.style={fill=white},
tick label style={font=\small},
label style={font=\small},
xmajorgrids,
ymajorgrids,
yminorgrids,
xminorticks=true, minor x tick num=4,
legend style={at={(0.01,0.99)}, anchor=north west},
legend style={legend cell align=left, align=left, draw=white!15!black,font=\small,nodes={scale=0.85, transform shape}}
]

\addplot [color=darkblue,solid,mark=o,line width=1.0pt]
table{%
	0	0.969000000000000
	0.250000000000000	0.914000000000000
	0.500000000000000	0.759000000000000
	0.750000000000000	0.548000000000000
	1	0.299700299700300
	1.25000000000000	0.141376060320452
	1.50000000000000	0.0442347390150398
	1.75000000000000	0.0100070049034324
	2	0.00175097276264591
};
\addlegendentry{$R0$}; 

\addplot [color=darkred,solid,mark=x,line width=1.0pt]
table{%
	0	0.972000000000000
	0.250000000000000	0.916000000000000
	0.500000000000000	0.760000000000000
	0.750000000000000	0.554000000000000
	1	0.309690309690310
	1.25000000000000	0.146559849198869
	1.50000000000000	0.0467413742258921
	1.75000000000000	0.0101404316354782
};
\addlegendentry{$R1$}; 

\addplot [color=lightgreen,solid,mark=triangle,line width=1.0pt]
table{%
	0	0.972000000000000
	0.250000000000000	0.916000000000000
	0.500000000000000	0.754000000000000
	0.750000000000000	0.538000000000000
	1	0.295704295704296
	1.25000000000000	0.130065975494816
	1.50000000000000	0.0429076968445886
	1.75000000000000	0.00890623436405484
};
\addlegendentry{$Rep$}; 

\addplot [color=darkgreen,solid,mark=square,line width=1.0pt]
table{%
	0	0.974000000000000
	0.250000000000000	0.927000000000000
	0.500000000000000	0.767000000000000
	0.750000000000000	0.575000000000000
	1	0.321678321678322
	1.25000000000000	0.148444863336475
	1.50000000000000	0.0520495429076968
	1.75000000000000	0.0109743487107642
	2	0.00245136186770428
};
\addlegendentry{$SPC$}; 

\addplot [color=orange,solid,mark=diamond,line width=1.0pt]
table{%
	0	0.970000000000000
	0.250000000000000	0.921000000000000
	0.500000000000000	0.761000000000000
	0.750000000000000	0.545000000000000
	1	0.295704295704296
	1.25000000000000	0.134307257304430
	1.50000000000000	0.0449719846652905
	1.75000000000000	0.00923980119416925
	2	0.00190661478599222
};
\addlegendentry{MSIB FSCL}; 

\addplot [color=black,solid,mark=+,line width=1.0pt]
table{%
	0	0.971000000000000
	0.250000000000000	0.918000000000000
	0.500000000000000	0.762000000000000
	0.750000000000000	0.554000000000000
	1	0.311688311688312
	1.25000000000000	0.144674835061263
	1.50000000000000	0.0468888233559422
	1.75000000000000	0.00997364822042096
	2	0.00180933852140078
};
\addlegendentry{MSIB SCL ap.};

\end{axis}
\end{tikzpicture}%
}}
	\end{minipage}
	\hspace{-15pt}
	\begin{minipage}{0.49\columnwidth}
		\centering		
		\subfloat[]{\label{fig:BLER_NodeEffect_MSIB}{\begin{tikzpicture}
\pgfplotsset{legend style={font=\small}}
\begin{axis}[%
width=0.65*\columnwidth,
height=1*\columnwidth,
scale only axis,
xmin=1.0,
xmax=2.25,
ymode=log,
ymin=8e-04,
ymax=1,
yminorticks=true,
ylabel={Block Error Rate},
xlabel={$E_b/N_0$ in dB},
axis background/.style={fill=white},
tick label style={font=\small},
label style={font=\small},
xmajorgrids,
ymajorgrids,
yminorgrids,
xminorticks=true, minor x tick num=4,
legend style={at={(0.01,0.99)}, anchor=north west},
legend style={legend cell align=left, align=left, draw=white!15!black,font=\small,nodes={scale=0.85, transform shape}}
]

\addplot [color=darkblue,solid,mark=o,line width=1.0pt]
table{%
	0	0.978000000000000
	0.250000000000000	0.908000000000000
	0.500000000000000	0.778000000000000
	0.750000000000000	0.565000000000000
	1	0.328000000000000
	1.25000000000000	0.140515222482436
	1.50000000000000	0.0419991600167997
	1.75000000000000	0.0101961050878564
	2	0.00184926060396851
};
\addlegendentry{$R0$}; 

\addplot [color=darkred,solid,mark=x,line width=1.0pt]
table{%
	0	0.977000000000000
	0.250000000000000	0.906000000000000
	0.500000000000000	0.780000000000000
	0.750000000000000	0.566000000000000
	1	0.320000000000000
	1.25000000000000	0.139578454332553
	1.50000000000000	0.0410191796164077
	1.75000000000000	0.0102300921048160
	2	0.00183693219994206
};
\addlegendentry{$R0,R1$}; 

\addplot [color=lightgreen,solid,mark=triangle,line width=1.0pt]
table{%
	0	0.976000000000000
	0.250000000000000	0.893000000000000
	0.500000000000000	0.771000000000000
	0.750000000000000	0.558000000000000
	1	0.309000000000000
	1.25000000000000	0.129274004683841
	1.50000000000000	0.0375192496150077
	1.75000000000000	0.00887061142643510
	2	0.00156570731136001
};
\addlegendentry{$R0,R1,Rep$}; 

\addplot [color=orange,solid,mark=diamond,line width=1.0pt]
table{%
	0	0.980000000000000
	0.250000000000000	0.898000000000000
	0.500000000000000	0.780000000000000
	0.750000000000000	0.567000000000000
	1	0.307000000000000
	1.25000000000000	0.139110070257611
	1.50000000000000	0.0404591908161837
	1.75000000000000	0.0100261700030588
	2	0.00184309640195528
};
\addlegendentry{MSIB FSCL}; 

\addplot [color=darkgreen,solid,mark=square,line width=1.0pt]
table{%
	0	0.977000000000000
	0.250000000000000	0.906000000000000
	0.500000000000000	0.788000000000000
	0.750000000000000	0.571000000000000
	1	0.324000000000000
	1.25000000000000	0.139110070257611
	1.50000000000000	0.0418591628167437
	1.75000000000000	0.0101281310539374
	2	0.00183693219994206
};
\addlegendentry{MSIB SCL ap.};

\end{axis}
\end{tikzpicture}%
}}
	\end{minipage}
	\caption{Effect of special nodes in MS\gls{ib} \gls{fssc}L decoding.}
	\label{fig:NodeEffect_MSIB}
		\vspace{-0.4cm}
\end{figure}
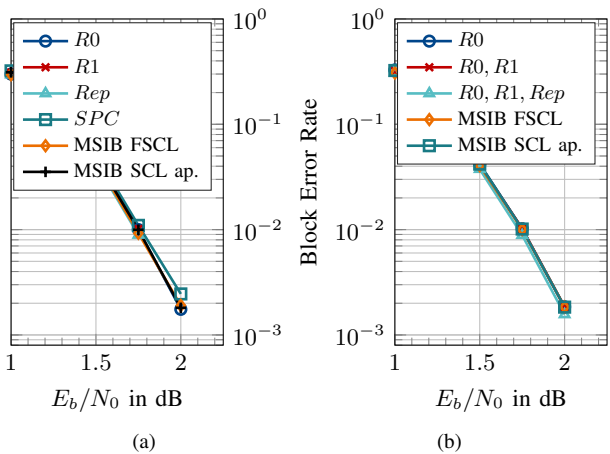
\begin{figure}[tb]
	\centering
	\begin{tikzpicture}
\begin{axis}[%
width=0.85*\columnwidth,
height=0.25*\columnwidth,
scale only axis,
xmin=0.5,
xmax=86.5,
ymin=0,
ymax=10,
y dir = reverse,
ylabel={depth $d$},
xlabel={$i_v$},
axis background/.style={fill=white},
tick label style={font=\small},
label style={font=\small},
xmajorgrids,
ymajorgrids,
yminorgrids,
legend columns=4, 
legend entries={{$R0$},{$R1$},{$Rep$},{$SPC$}},
legend style={at={(0.5,0.99)}, anchor=north,font=\small,nodes={scale=0.9, transform shape}},
]
\addlegendimage{line width=3.0pt,darkblue}
\addlegendimage{line width=3.0pt,darkred}
\addlegendimage{line width=3.0pt,lightgreen}
\addlegendimage{line width=3.0pt,orange}
\addplot [ycomb, color=darkblue,line width=2.0pt]
table{%
2	5
3	6
4	7
5	8
6	9
8	6
9	7
10	8
18	5
19	6
20	7
33	9
36	8
51	8
52	9
59	7
80	9
};

\addplot [ycomb, color=darkred,line width=2.0pt]
table{%
7	9
17	7
29	8
34	9
35	8
44	5
53	9
68	6
70	8
71	7
72	6
73	5
79	7
81	9
82	8
83	7
84	6
};

\addplot [ycomb, color=lightgreen,line width=2.0pt]
table{%
	1	3
	12	7
	13	8
	15	8
	21	8
	23	6
	24	7
	25	8
	27	7
	28	8
	31	6
	32	7
	40	8
	45	5
	46	6
	47	7
	48	8
	50	6
	55	8
	60	8
	62	8
	65	8
	69	8
	74	7
	75	8
	77	8
};

\addplot [ycomb, color=orange,line width=2.0pt]
table{%
	11	8
	14	8
	16	8
	22	8
	26	8
	30	6
	37	8
	38	7
	39	6
	41	8
	42	7
	43	6
	49	8
	54	7
	56	8
	57	7
	58	6
	61	8
	63	8
	64	7
	66	8
	67	7
	76	8
	78	8
	85	4
	86	3
};

\end{axis}
\end{tikzpicture}%
	\vspace{-0.3cm}
	\caption{Sequence of nodes and their depth $d$ in the decoder tree under FSC schedule for $N{=}1024, K{=}512$ and $N_{\text{CRC}}{=}16$. i.e., $|\mset{A}|=528$.}
	\label{fig:FSSCL_schedule_depth}
\end{figure}
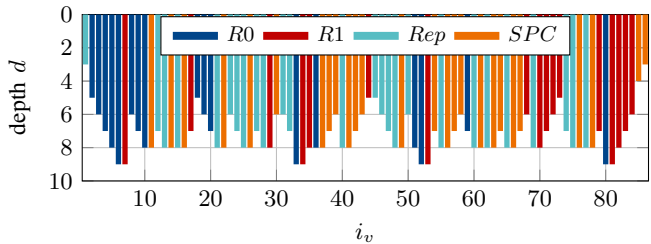

Table.~\ref{tab:lut_number_data} presents the number of unique \glspl{lut} required by the proposed decoders for various code rates.
The number of decoding tables required for \gls{lut}-based \gls{scl} decoding is equal to the number of edges in its decoder tree. The number of translation table is equal to the number of leaf nodes in the decoder tree.
An \gls{ib} \gls{scl} decoder requires $2N{-}2{=}2046$ distinct decoding tables and $N{=}1024$ translation tables\cite{shah_design_2019} regardless of the code rate. 
The MSIB SCL \cite{shah_MSIB_2023} requires $N{=}1024$ decoding as well as $N$ translation tables.
In comparison, the IB and MSIB FSCL decoders require as few as $71$ translation tables. Depending upon the code rate, the proposed IB FSCL requires as few as $140$ decoding tables instead of $2046$. Similarly, the MSIB FSCL requires $71$ to $86$ decoding tables. These number show a remarkable  reduction in the number of  \glspl{lut} used for performing the proposed finite alphabet decoding, i.e., up to $93\%$.

 \section{Conclusion}

In this paper, the design of \gls{lut}-based quantized polar decoders was extended to use fast decoding schedules. 
The fast schedules increase the decoding speed of \gls{scl} decoders by deploying special constituent decoders. 
It was shown that the \glspl{lut} designed for conventional \gls{sc} schedule can be used for the fast \gls{sc} schedule without needing any change in the design process of the \glspl{lut}. 
The potential increase in decoding speed of the fast \gls{sc} schedules  has  negligible effect on the error correction performance. 
Most importantly, the need for distinct \glspl{lut} in the proposed finite alphabet decoders reduces by up to $93\%$. 

\begin{table}[tb]
	\renewcommand{\arraystretch}{1.3}
	\setlength\tabcolsep{2.5pt}
	\caption{number of required lookup tables for proposed decoders}
	\vspace{-15pt}
	\begin{center}
		\begin{tabular}{|c|c|c|c|c|c|c|}	
			\hline
			& \multicolumn{3}{c|}{IB FSCL} & \multicolumn{3}{c|}{MSIB FSCL}\\
			\textbf{table type}	&$R{=}0.25$ &$R{=}0.5$&$R{=}0.25$&$R{=}0.25$  &$R{=}0.5$&$R{=}0.25$ \\
			\hline
			Decoding  	&140	&170			&148	&71		&86	&75 \\
			\hline
			Translation     	&71	&86  	  &75	&71		&86 	&75 \\
			\hline
		\end{tabular}
		\label{tab:lut_number_data}
	\end{center}
	\vspace{-10pt}
\end{table}

\bibliographystyle{IEEEtran}
\bibliography{literature}

\end{document}